\documentclass[a4paper,11pt]{article}
\pdfoutput=1 

\usepackage{jheppub} 

\usepackage[T1]{fontenc} 

\usepackage{bm}
\newcommand{\beq}{\begin{equation}}
\newcommand{\eeq}{\end{equation}}
\title{\huge{\textbf{From Spin Vertex to String Vertex}}}

\author{Yunfeng Jiang}
\author{and Andrei Petrovskii}
\emailAdd{jinagyf2008@gmail.com
, andreypetrovskij@gmail.com}
\affiliation{Institut de Physique Th\'eorique,
DSM, CEA, URA2306 CNRS\\Saclay, F-91191 Gif-sur-Yvette,
France}

\abstract{In the recent publication \cite{Jiang:SpinVertex} the spin vertex was introduced as a new approach for computing three-point functions in $\mathcal{N}=4$ SYM. In this note we consider the BMN limit of the spin vertex for scalar excitations and show that it reproduces the string vertex in the light-cone string field theory which describes the string interactions in the pp-wave background at the leading order of $\lambda'$ expansion. This is achieved by introducing a polynomial representation for the spin vertex. We derive the Neumann coefficients from the spin vertex at weak coupling and show they match with the Neumann coefficients from the string field theory.
}

\begin{document}
\maketitle
\flushbottom

\section{Introduction}
One of the most important progress in theoretical physics in the last two decades is the discovery of the AdS/CFT correspondence \cite{Maldecena:AdS-CFT,Witten:1998,GKP:1998}, which reveals the duality between gauge theories in flat spacetime and gravity theories in higher dimensions. This duality is a strong/weak duality, namely the weak coupling limit of one theory is dual to the strong coupling limit of the other theory and vice versa. On the one hand, this fact creates the exciting possibility of tackling gauge theories at strong coupling. On the other hand, this also makes it notoriously hard to check the duality in full generality. Early checks of this correspondence mainly concentrate on BPS objects which are protected by symmetry and do not depend on the coupling constant at all. An important breakthrough was made by Berenstein, Maldacena and Nastate \cite{BMN} who suggested to study a specific sector of strongly coupled ${\cal N}=4$ SYM theory which contains operators of infinite length and finite number of excitations, such as
\beq
\mathrm{Tr}( ... ZZZ \phi_i ZZ... ZZ\psi_j ZZZ...)
\eeq
where the complex scalar field $Z$ is regarded as the `vacuum' and the other fields are regarded as `excitations'. This sector is the so-called BMN limit of the $\mathcal{N}=4$ SYM theory. In the BMN limit, both the scaling dimension $\Delta$ and the $R$-charge $J$ of an operator are divergent, but their difference $\Delta-J$ remains finite and allows a perturbative computation in terms of an effective coupling constant $\lambda' = \frac{\lambda}{J^2}$. The BMN limit provides a regime where one can compare the quantities from both sides of duality directly. Nevertheless it's been shown that gauge and string theory computations in BMN limit can have some deviations at loop order, which are generally explained by the argument that in gauge and string theories the limits $J\to \infty$, $\lambda' \to 0$ are taken in different orders. \par

From string theory side the BMN limit is obtained by taking the Penrose limit of the $AdS_5\times S^5$ string sigma model \cite{Papadopoulos:pp-wave}. The metric takes the form
\beq
ds^2 = -4dX^{+}dX^{-} - \mu^2 X^2 (dX^{+})^2+dX_{i}dX^{i},
\eeq
and describes a point-like string sitting at the center of $AdS_5$ and moving close to the speed of light along the equator of $S^5$. The parameter $\mu$ has dimension of mass and putting it to zero corresponds to taking flat space limit. Although the pp-wave background is curved the quantization in light-cone gauge can be performed readily.\par

The holographic dictionary between parameters is given by the following relations:
\beq
\lambda' = \frac{\lambda}{J^2} = \frac{1}{(\mu p^{+}\alpha')^2},
\quad \frac{J^2}{N} = 4\pi g_s (\mu p^{+}\alpha')^2.
\eeq

The spectral problem of $\mathcal{N}=4$ SYM theory is now considered to be solved, at least conceptually. The natural next step is to compute the three-point functions, which contains the dynamical information of the theory. In recently years, substantial progress have been made both at strong coupling \cite{Zarembo:3pt1,Janik:3pt1,Janik:3pt2,Janik:HHL,Costa:3pt,Komatsu:3pt1,Komatsu:3pt2,Komatsu:su2} and weak coupling \cite{EGSV,EGSV:Tailoring2,EGSV:Tailoring3,Foda:3ptdeterminant,Foda:su3,GV,GV:quantumintegrability,Jiang:OneLoop} from integrability. The most important problem is to identify common integrable structures and look for a framework which is applicable at any coupling. In general, it is very hard to construct such a framework. However, it the much simpler BMN limit, this seems feasible and its better understanding might shed some light on the general case.

The string interactions in the pp-wave geometry are described by the light-cone string field theory (SFT). The central object of SFT is the cubic \emph{string vertex} which is a special entangled state living in the three-string Hilbert space. It contains all information of string interactions. In our recent publication \cite{Jiang:SpinVertex} (see also \cite{Kazama:Vertex}), we constructed a weak coupling counterpart of the string vertex, which is termed the \emph{spin vertex}. The spin vertex also contains all information about operator/spin chain interactions. We constructed the spin vertex explicitly at tree level. While the string vertex is applicable in the BMN limit, the spin vertex works for general spin chain states at leading order. In this work, we examine the BMN limit of the spin vertex and show that it reproduces the string vertex in SFT at the leading order of $\lambda'$ expansion.\par

The rest of the paper is organized as follows. In section 2, we give a brief review of the light-cone string field theory and the string vertex. In particular, we will focus on the proposal by Dobashi and Yoneya \cite{Dobashi:Resolving,Dobashi:3pt1} (see also \cite{LeeRusso}) since their proposal works not only for the extremal correlation functions but also non-extremal correlation functions. At weak coupling, it is much easier to consider the non-extremal case since there is no mixture between single trace and double trace operators. In section 3, we shall provide a polynomial representation for the spin vertex in the pure scalar sector, which allows taking the BMN limit straightforwardly and comparing to SFT. After these preparations, we show how to reproduce the string vertex from spin vertex at leading order of $\lambda'$ expansion in section 4. Finally, we conclude and present further directions to pursue in section 5.

\section{A Brief Review of Light-Cone String Field Theory}
In this section we review briefly the light-cone string field theory for strings on the pp-wave background \cite{Volovich:PPwave1,Volovich:PPwave2,Pankiewicz:Morecomment,PankiewiczStephanski,Pankiewicz:Alternative} and refer the interested readers to \cite{Plefka:BMNlecture,Pankiewicz:Thesis} and references therein for more detail.\par
The fundamental object in SFT is the string field operator $\Phi$
which creates and destroys strings. The Hilbert space on which the string field operator acts is a direct sum of $n$-string Hilbert spaces:
\beq
\begin{split}
&{\cal H} = |\text{vac}\rangle \oplus {\cal H}_1 \oplus{\cal H}_2 \oplus..., \hskip 10 pt
{\cal H}_n = \underbrace{{\cal H}_1\otimes ... \otimes {\cal H}_1}_{n}\\
\end{split}
\eeq
where ${\cal H}_1$ is the Hilbert space of a single string. The string interactions are described by the matrix elements of the Hamiltonian which has the following expansion in coupling constant $g_s$
\beq
H = H_2 + g_s H_3 + g_s^2 H_4+...,
\eeq
where $H_2$ is the free part of the Hamiltonian and $H_k$ describes interactions involving $k$ strings. In this paper, we focus on the cubic interactions which are described by the following matrix elements
\beq
\lambda_{123} = \langle 2| \langle 3| H_3 | 1\rangle = \langle 1|
\langle 2| \langle 3|H_3\rangle,
\eeq
where $\langle1|,\langle2|,\langle3|$ are three string states. In the second equality the matrix element is written in a more customary way by introducing the so-called cubic \emph{string vertex} $|H_3\rangle$.

The principle to construct the string vertex is simple. For bosonic strings, the string vertex $|H_3\rangle$ is constructed by requiring worldsheet continuity at interacting point, which can be realized by imposing a delta functional $\Delta\Big(X_1(\sigma)-X_2(\sigma)-X_3(\sigma)\Big)$ in the functional integral over all possible configurations of three strings. The integral can be computed straightforwardly, leading to the following form of the bosonic string vertex \footnote{Note that we use different notations from the ones in \cite{Dobashi:3pt1}. Our creation operator $a_{m}^{(r)i\dagger}$ is denoted by $\alpha_m^{(r)i\dagger}$ in \cite{Dobashi:3pt1} and our Neumann coefficient $N_{mn}^{rs}$ is denoted by $\tilde{N}_{mn}^{rs}$ in \cite{Dobashi:3pt1}.}
\beq
\label{overlap}
|V\rangle=\exp\Bigg(-\frac{1}{2}\sum_{m,n=-\infty}^\infty\sum\limits_{r,s=1}^3\sum_{i=1}^8 a_m^{(r)i\dagger}N_{mn}^{rs}a_n^{(s)i\dagger}\Bigg)
|0\rangle\!\rangle.
\eeq
Here $|0\rangle\!\rangle$ denotes the vacuum of three-string Hilbert space $|0\rangle\!\rangle\equiv|0\rangle_1\otimes|0\rangle_2\otimes|0\rangle_3$. The indices $r$ and $s$ denote the number of strings, $i$ denotes the polarization of the excitation and $m,n$ are the mode numbers of the excitations. The quantities $N_{mn}^{rs}$ are called Neumann coefficients and characterize the interactions between excitations of different strings.

For superstrings, in addition to worldsheet continuity, one also needs to require that supersymmetry is respected by the string vertex. This can be achieved by acting a new operator ${\cal P}$ on the exponential part (\ref{overlap}). This operator can be written as a quadratic polynomial of creation operators $a_m^{(r)i}$ (bosonic as well as fermionic, but since we consider the bosonic sector, we will drop the fermionic part) and is called the \emph{prefactor}. The string vertex for superstring thus takes the following form
\beq
|H_3\rangle = {\cal P}|V\rangle.
\eeq
However, it turns out that supersymmetry is not restrictive enough to fix the prefactor uniquely and there have been several proposals in the literature originating from different motivations. In this paper, we will use the prefactor proposal by Dobashi and Yoneya \cite{Dobashi:Resolving}.
The reason is that their proposal has the virtue that works for both extremal and non-extremal\footnote{The extremal and non-extremal correlation functions corresponds to the impurity preserving and impurity non-preserving processes in \cite{Dobashi:3pt1} respectively.} correlation functions \cite{Dobashi:3pt1}. Interestingly, the prefactor of Dobashi and Yoneya is the half sum of two prefactors ${\cal P}_{1}$ and ${\cal P}_{2}$ proposed in \cite{Volovich:PPwave1,Volovich:PPwave2,Pankiewicz:Morecomment,PankiewiczStephanski} and \cite{DiVecchia:SFT2} respectively
\beq
{\cal P}_{h} = \frac{1}{2}{\cal P}_{1} + \frac{1}{2} {\cal P}_{2},
\eeq
more explicitly
\beq
{\cal P}_{h} = \sum\limits_{r=1}^3\Bigg(\sum\limits_{i=5}^{8}\sum\limits_{m=0}^{\infty} \frac{\omega_{m}^{(r)}}{\alpha_{(r)}}a_{m}^{(r)i\dagger} a_{m}^{(r)i}+\sum\limits_{i=1}^{4}\sum\limits_{m=0}^{\infty} \frac{\omega_{m}^{(r)}}{ \alpha_{(r)}} a_{-m}^{i(r)\dagger} a_{-m}^{(r)i}\Bigg),
\eeq
where $\omega_{n}^{(r)} = \sqrt{n^2+\mu^2\alpha^2_{(r)}}$ and
$\alpha_{(r)} = \alpha' p^{+}_{(r)}$.

After one fixes the string vertex, the matrix elements of $H_3$ can be computed straightforwardly. According to \cite{Dobashi:Resolving}, the holographic relation between matrix element of the $H_3$ and OPE coefficients in BMN limit is given by
\begin{align}
\label{HolRe2}
C_{123}=\frac{\sqrt{J_1J_2J_3}}{N}\frac{
G(\Delta_1,\Delta_2,\Delta_3)}{\mu(\Delta_2+\Delta_3-\Delta_1)}
\lambda_{123}.
\end{align}
Here $C_{123}$ is the structure constant of the three-point correlation function
\begin{align}
\label{BigC}
\langle\mathcal{O}_1(x_1)\mathcal{O}_2(x_2)\mathcal{O}_3(x_3)
\rangle=\frac{C_{123}}{|x_{12}|^{\Delta_1+\Delta_2-\Delta_3}
|x_{13}|^{\Delta_1+\Delta_3-\Delta_2}|x_{23}|^{\Delta_2+\Delta_3-
\Delta_1}}
\end{align}
where $x^\mu_{ij}=x^\mu_i-x^\mu_j$, $\Delta_i$ is the conformal dimension of the operator ${\cal O}_i$ and the function $G(\Delta_1,\Delta_2,\Delta_3)$ reads
\begin{align}
\label{G}
G(\Delta_1,\Delta_2,\Delta_3)=
\left(f\frac{J_2J_3}{J_1}\right)^{-(\Delta_2+\Delta_3-\Delta_1)/2}
\Gamma\left(\frac{\Delta_2+\Delta_3-\Delta_1}{2}+1\right),
\end{align}
where the function $f$ is defined in \cite{Dobashi:3pt1}. Finally, we want to emphasis that the holographic relation between the matrix elements of $H_3$ and the OPE coefficient in $\mathcal{N}=4$ is not completely understood. The holographic relation (\ref{HolRe2}) works well at the leading order \cite{Dobashi:3pt1,Schulgin:BMN}. However, at higher loop order, the large $\mu$ expansion of the function $G(\Delta_1,\Delta_2,\Delta_3)$ give rises to non-perturbative terms such as $\log\mu$, the interpretation of which is still unclear.

\section{Polynomial Representation of Spin Vertex}
The notion of spin vertex was introduced in \cite{Alday:Vertex,Jiang:SpinVertex} where the spin vertex at tree level was constructed in terms of a set of bosonic and fermionic oscillators. In this paper we will the compact sector of $\mathcal{N}=4$ SYM, which is built from six fundamental scalar fields: $X$, $\bar X$, $Y$, $\bar Y$, $Z$, $\bar Z$ and forms a representation of $SO(6)$ group. The spin vertex for the compact sector is constructed by \emph{fermionic oscillators}. On the other hand, the cubic string vertex in SFT is constructed in terms of \emph{bosonic oscillators}. In order to derive SFT from spin vertex, we introduce a polynomial representation for the spin vertex in the compact sector. From the polynomial representation, it is then straightforward to construct the spin vertex using bosonic oscillators and make contact with SFT.\par

\begin{figure}[h!]
\begin{center}
\includegraphics[scale=0.5]{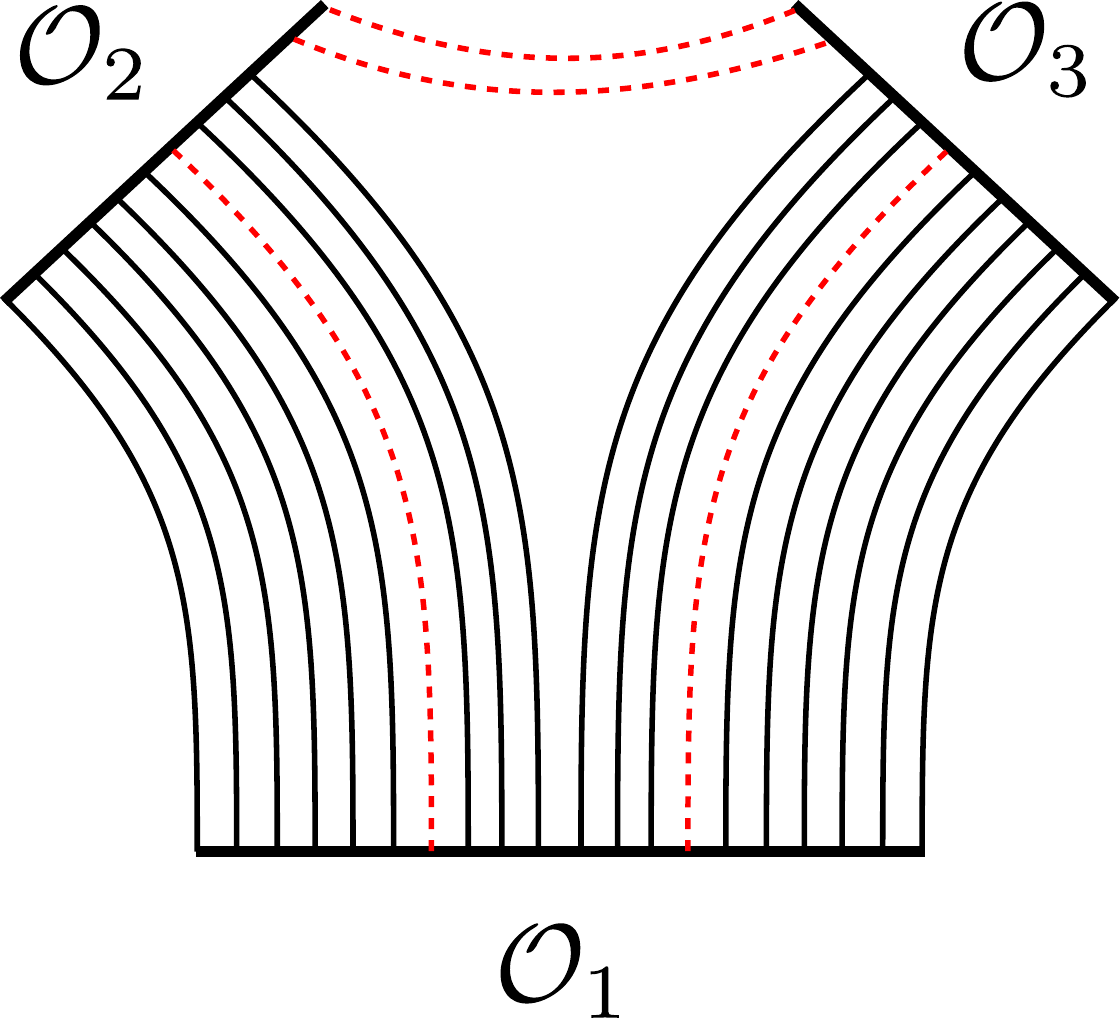}
\caption{The configuration for three-point functions. The black lines correspond to the contractions of $Z$ and $\bar{Z}$ and the red dashed lines correspond to the contractions of excitations $\phi^i$, $i=1,2,3,4$. }
\label{BMN}
\end{center}
\end{figure}

In what follows, as a convention, we choose operator $\mathcal{O}_1$ to be the ``incoming'' state and $\mathcal{O}_2$, $\mathcal{O}_3$ to be ``outgoing'' states, as is depicted in Fig\ref{BMN}.
The BMN vacuum for the incoming and outgoing states at each site are a scalar field $Z$ and $\bar{Z}$ respectively, where
\begin{align}
Z=\frac{1}{\sqrt{2}}(\phi^5+i\phi^6),\quad \bar{Z}=\frac{1}{\sqrt{2}}(\phi^5-i\phi^6)
\end{align}
The ``vacuum'' can have different excitations, namely scalars excitations, vector excitations and fermionic excitations. As mentioned before, we consider here only scalar excitations. The three operators are thus made of the following fields
\begin{align}
\label{setup}
\mathcal{O}_1:\{Z,\phi^i\},\quad \mathcal{O}_2:\{\bar{Z},\phi^i\},\quad \mathcal{O}_3:\{\bar{Z},\phi^i\},\quad i=1,2,3,4
\end{align}
Following \cite{Dobashi:3pt1}, we normalize the operators as
\begin{align}
\langle\overline{\mathcal{O}}_i(x_1)\mathcal{O}_i(x_2)\rangle=\frac{1}{(x_1-x_2)^{2\Delta_i}},\quad i=1,2,3
\end{align}
where $\Delta_i$ is the scaling dimension of the $i$-th operator and at leading order we have $\Delta_i=L_i$. The three-point function of three scalar operators is determined up to a constant called \emph{structure constant} by conformal symmetry
\begin{align}
\langle\mathcal{O}_1(x_1)\mathcal{O}_2(x_2)\mathcal{O}_3(x_3)\rangle=\frac{\sqrt{L_1L_2L_3}}{N}\frac{c_{123}}
{|x_{12}|^{\Delta_1+\Delta_2-\Delta_3}|x_{13}|^{\Delta_1+\Delta_3-\Delta_2}|x_{23}|^{\Delta_2+\Delta_3-\Delta_1}}.
\end{align}
At the leading order, we have
\begin{align}
\Delta_i+\Delta_j-\Delta_k=L_i+L_j-L_k=2L_{ij},\quad i,j,k=1,2,3
\end{align}
where $L_{ij}$ is the number of propagators between operators $i$ and $j$. The structure constant $c_{123}$ is related to the one in (\ref{BigC}) by $C_{123}=\frac{\sqrt{L_1L_2L_3}}{N}c_{123}$ . The computation of $c_{123}$ can be formulated in the spin vertex formalism.
In our case, we only need the scalar sector of the spin vertex, which can be obtained by a truncation from the full spin vertex \cite{Jiang:SpinVertex}. The scalar sector corresponds to the fermionic part of the spin vertex.
Let us recall the fermionic part of the two-point spin vertex at each site from \cite{Jiang:SpinVertex}
\begin{align}
\label{vertex}
|v_{12}\rangle=\exp\sum_{i=1,2}\left(d_{i}^{(1)\dagger}d_{i}^{(2)}+c_{i}^{(1)\dagger}c_{i}^{(2)}\right)|0\rangle^{(1)}\otimes|\bar{0}\rangle^{(2)}
\end{align}
where $c_i,c_i^\dagger$ and $d_i,d_i^\dagger$ are two sets of fermionic oscillators, satisfying the usual anti-commutation relations
\begin{align}
\{c_i,c_j^\dagger\}=\delta_{ij},\quad \{d_i,d_j^\dagger\}=\delta_{ij}
\end{align}
The rest anti-commutation relations vanish. The vacuum $|0\rangle$ and dual vacuum $|\bar{0}\rangle$ correspond to the scalar field $Z$ and $\bar{Z}$ respectively and satisfy
\begin{align}
c_i|0\rangle=d_i|0\rangle=0,\quad c_i^\dagger|\bar{0}\rangle=d_i^\dagger|\bar{0}\rangle=0,\quad  i=1,2.
\end{align}
More explicitly, we can write $|\bar{0}\rangle=c_1^\dagger c_2^\dagger d_1^\dagger d_2^\dagger|0\rangle$. We want to expand (\ref{vertex}) and write the spin vertex in a more transparent way. The expansion leads to a sum of many terms, among which we keep only the terms with $N_c=N_d$, where $N_c$ and $N_d$ counts the number of operators $c_i,c_i^\dagger$ and $d_i,d_i^\dagger$, respectively. The reason is that only the terms satisfying $N_c=N_d$ corresponds to scalar fields. The terms which satisfy this condition in the expansion read
\begin{align}
|v_{12}\rangle_{\text{scalar}}\equiv|v_{12}\rangle_{N_c=N_d}=|0\rangle\otimes|\bar{0}\rangle+\sum_{i,j=1}^2 d_i^{(1)\dagger}c_j^{(1)\dagger}|0\rangle\otimes d_i^{(2)}c_j^{(2)}|\bar{0}\rangle+|\bar{0}\rangle\otimes|0\rangle
\end{align}
Let us define the following states corresponding to the scalar fields $X,Y$ and their hermitian conjugates $\bar{X}$, $\bar{Y}$
\begin{align}
&|X\rangle\equiv d_1^\dagger c_1^\dagger|0\rangle=d_2c_2|\bar{0}\rangle,\quad |\bar{X}\rangle\equiv d_2^\dagger c_2^\dagger|0\rangle=d_1c_1|\bar{0}\rangle\\\nonumber
&|Y\rangle\equiv d_1^\dagger c_2^\dagger|0\rangle=c_1d_2|\bar{0}\rangle,\quad |\bar{Y}\rangle\equiv c_1^\dagger d_2^\dagger|0\rangle=d_1c_2|\bar{0}\rangle.
\end{align}
Then we have
\begin{align}
|v_{12}\rangle_{\text{scalar}}=
\sum_{a=X,Y,Z}\left(|a\rangle\otimes|\bar{a}\rangle+
|\bar{a}\rangle\otimes|a\rangle\right),
\end{align}
We can further define the states corresponding to the real scalar fields $\phi_i$, $i=1,2,3,4$ as the following
\begin{align}
&|X\rangle=\frac{1}{\sqrt{2}}(|\phi_1\rangle+i|\phi_2\rangle),\quad |\bar{X}\rangle=\frac{1}{\sqrt{2}}(|\phi_1\rangle-i|\phi_2\rangle) \\\nonumber
&|Y\rangle=\frac{1}{\sqrt{2}}(|\phi_3\rangle+i|\phi_4\rangle),\quad |\bar{Y}\rangle=\frac{1}{\sqrt{2}}(|\phi_3\rangle-i|\phi_4\rangle)
\end{align}
which gives
\begin{align}
\label{vv}
|v_{12}\rangle_{\text{scalar}}=|Z\rangle\otimes|\bar{Z}\rangle+|\bar{Z}\rangle\otimes|Z\rangle+\sum_{i=1}^4|\phi_i\rangle\otimes|\phi_i\rangle
\end{align}
The spin vertex for three-point functions of our set-up is obtained by taking the appropriate tensor products of the vertex in (\ref{vv}), which reads
\begin{align}
\label{Vf}
|V_3\rangle=&\prod_{k=1}^{L_{12}}\left(|Z\rangle_{L_1-k+1}^{(1)}
\otimes|\bar{Z}\rangle_k^{(2)}+
|\bar{Z}\rangle_{L_1-k+1}^{(1)}\otimes|Z\rangle_k^{(2)}+
\sum_{i=1}^4|\phi_i\rangle^{(1)}_{L_1-k+1}\otimes|\phi_i\rangle_k^{(2)}   \right)\otimes\\\nonumber
&\prod_{k=1}^{L_{13}}\left(|Z\rangle_k^{(1)}\otimes|\bar{Z}\rangle_{L_3-k+1}^{(3)}+|\bar{Z}\rangle_k^{(1)}\otimes|Z\rangle_{L_3-k+1}^{(3)}+\sum_{i=1}^4|\phi_i\rangle^{(1)}_k\otimes|\phi_i\rangle_{L_3-k+1}^{(3)}   \right)\otimes\\\nonumber
&\prod_{k=1}^{L_{23}}\left( |Z\rangle_{L_2-k+1}^{(2)}\otimes|\bar{Z}\rangle_k^{(3)}+|\bar{Z}\rangle_{L_2-k+1}^{(2)}\otimes|Z\rangle_k^{(3)}+\sum_{i=1}^4|\phi_i\rangle^{(2)}_{L_2-k+1}\otimes|\phi_i\rangle_{k}^{(3)}  \right).
\end{align}
where the upper indices denote the spin chains and the lower indices denote the positions on the corresponding spin chains. In order to obtain a polynomial representation, let us define the following generating states for the three spin chains
\begin{align}
&|F_1\rangle=\prod_{k=1}^{L_1}\left(|Z\rangle_k+\sum_{i=1}^4 x^i_k|\phi^i\rangle_k\right)\\\nonumber
&|F_2\rangle=\prod_{k=1}^{L_2}\left(|\bar{Z}\rangle_k+\sum_{i=1}^4 y^i_k|\phi^i\rangle_k\right)\\\nonumber
&|F_3\rangle=\prod_{k=1}^{L_3}\left(|\bar{Z}\rangle_k+\sum_{i=1}^4 z^i_k|\phi^i\rangle_k\right)
\end{align}
where we couple each excitation with an auxiliary variable. The three states can be obtained in the following way
\begin{align}
|1\rangle=\left.\Psi_1(\bm{\partial_x})|F_1\rangle\right|_{\bm{x}^i=0},\quad |2\rangle=\left.\Psi_2(\bm{\partial_y})|F_2\rangle\right|_{\bm{y}^i=0},\quad |3\rangle=\left.\Psi_3(\bm{\partial_z})|F_3\rangle\right|_{\bm{z}^i=0}
\end{align}
where $\Psi_1(\bm{\partial_x}),\Psi_2(\bm{\partial_y})$ and $\Psi_3(\bm{\partial_z})$ are three differential operators of the following form
\begin{align}
\label{differential}
\Psi(\bm{\partial_x})=\sum_{\bm{n},I}c^I_{\bm{n}}\prod_{k=1}^L(\partial^{i_k}_k)^{n_k},\qquad \partial_k^{i_k}\equiv\frac{\partial}{\partial x^{i_k}_k},\quad n_k=0,1
\end{align}
Here $\bm{n}=\{n_1,\cdots,n_L\}$ and $I$ is a collective index indicating the polarizations of the excitations. Each differential operator specifies one spin chain state. Integrability of the spin chain usually provides systematic way to construct the differential operators. In our case, for example, the differential operator can be constructed systematically by the nested Bethe ansatz.
According the the spin vertex formalism
\begin{align}
c_{123}=&\langle1|\langle2|\langle3|V_3\rangle\\\nonumber
=&\left.\Psi_1(\bm{\partial_x})\Psi_2(\bm{\partial_y})\Psi_3(\bm{\partial_z})\langle F_1|\langle F_2|\langle F_3|V_3\rangle\right|_{\bm{x}^i,\bm{y}^i,\bm{z}^i=0}\\\nonumber
=&\left.\Psi_1(\bm{\partial_x})\Psi_2(\bm{\partial_y})\Psi_3(\bm{\partial_z})V_3(\bm{x},\bm{y},\bm{z})\right|_{\bm{x}^i,\bm{y}^i,\bm{z}^i=0}
\end{align}
where $V_3(\bm{x}^i,\bm{y}^i,\bm{z}^i)$ is the polynomial representation of the spin vertex which reads
\begin{align}
\label{vertexso4}
V_3(\bm{x^i},\bm{y^i},\bm{z^i})=\prod_{k=1}^{L_{12}}(1+y^i_k x_{L_1-k+1}^i)\prod_{k=1}^{L_{13}}(1+x_k^i z^i_{L_3-k+1})\prod_{k=1}^{L_{23}}z_k^i y_{L_2-k+1}^i
\end{align}
In the polynomial representation, the spin vertex (\ref{vertexso4}) is given by a polynomial in terms auxiliary variables $\bm{x}^i,\bm{y}^i,\bm{z}^i$. The states are given by three differential operators, which can be constructed by Bethe ansatz. Note that the commutation relations of $\partial/\partial x$ and $x$ are the same as commutation relations of bosonic creation and annihilation operators. Therefore, we can map the auxiliary variables and the corresponding derivatives into creation and annihilation operators
\begin{align}
\label{map}
&x_k^i\rightarrow \alpha^{(1)i\dagger}_k,\quad y_k^i\rightarrow \alpha^{(2)i\dagger}_k,\quad z_k^i\rightarrow \alpha^{(3)i\dagger}_k\\\nonumber
&\frac{\partial}{\partial x_k^i}\rightarrow \alpha^{(1)i}_k,\quad \frac{\partial}{\partial y_k^i}\rightarrow \alpha^{(2)i}_k,\quad \frac{\partial}{\partial z_k^i}\rightarrow \alpha^{(3)i}_k
\end{align}
and the spin vertex can be written as
\begin{align}
\label{V3a}
|V_3\rangle_B\equiv V_3\left(\alpha^{(1)i\dagger}_k,\alpha^{(2)i\dagger}_k,\alpha^{(3)i\dagger}_k\right)|0\rangle_B
\end{align}
here we use $|V_3\rangle_B$ to emphasis that this vertex is constructed by \emph{bosonic} oscillators instead of \emph{fermionic} ones as in (\ref{Vf}). The new Fock vacuum is defined to be the state that is annihilated by all the bosonic annihilation operators
\begin{align}
\alpha^{(r)i}_k|0\rangle_B=0
\end{align}
The corresponding states can be written as
\begin{align}
\langle1|\equiv \langle0|\Psi_1\left(\alpha^{(1)i}_k\right),\quad \langle2|\equiv \langle0|\Psi_2\left(\alpha^{(2)i}_k\right),\quad \langle3|\equiv \langle0|\Psi_3\left(\alpha^{(3)i}_k\right)
\end{align}
and the structure constant is given by
\begin{align}
c_{123}=\langle 1|\langle 2|\langle 3|V_3\rangle_B.
\end{align}
where both the spin vertex and the states are now constructed by bosonic oscillators. The formulation we described up to now is applicable to any three states in our set-up. In the next section, we will take the BMN limit of (\ref{V3a}) and show that it reproduces the string vertex in SFT.

\section{BMN Limit of the Spin Vertex}
The BMN limit for $\mathcal{N}=4$ SYM mainly contains two approximations. The first is a dilute gas approximation, which means the number of excitations is finite and are distributed sparsely. The second approximation is that the momenta of the excitations are small and scales like $\sim 1/L$ where $L$ is the length of the spin chain, which is taken to be very large. \par
We start by recalling the spin vertex in polynomial representation in (\ref{vertexso4})
\begin{align}
V_3(\bm{x}^i,\bm{y}^i,\bm{z}^i)=\prod_{k=1}^{L_{12}}(1+x_{L_1-k+1}^i y_k^i)\prod_{k=1}^{L_{13}}(1+z_{L_3-k+1}^i x_k^i)\prod_{k=1}^{L_{23}} y_{L_2-k+1}^i z_k^i
\end{align}
The length of each spin chain is equal to the sum of number of vacuum field $Z$ or $\bar{Z}$ and number of excitations. We denote the number of vacuum fields by $J_i$ and the number of excitations by $N_i$ so that the length of the spin chain $L_i$ is given by $L_i=J_i+N_i$. By dilute gas assumption, we have $J_i\gg N_i$ and hence $L_i\simeq J_i$. Due to charge conservation, we have $J_1=J_2+J_3$, hence we can write $J_1 = J$, $J_2= rJ$, $J_3=(1-r)J$, where $0<r<1$.
The number of contractions between different operators are approximately
\begin{align}
L_{12}\simeq J_2= rJ,\quad L_{13}\simeq J_3=(1-r)J,\quad L_{23}=M=\frac{1}{2}(N_2+N_3-N_1).
\end{align}
In what follows, we use $M$ to denote the number of contractions between the two `out-going' operators. By BMN assumption, $M\ll J$. The three-point functions for $M=0$ are called \emph{impurity preserving}, or \emph{extremal} while for $M\ne 0$ are called \emph{impurity non-preserving} or \emph{non-extremal}. For the extremal correlator, when diagonalizing anomalous dimension matrix one have to deal with the mixing between single trace and double trace operators. On the contrary, for the non-extremal cases, the contribution from double trace operators are $1/N$-suppressed and can be neglected in the planar limit, which makes the computation much simpler from gauge theory aspect. On the other hand,  the earlier proposals for string field theory and duality relations work only for the extremal cases. A string field theory applicable to non-extremal cases as well as extremal is the holographic string field theory proposed by Dobashi and Yoneya \cite{Dobashi:Resolving,Dobashi:3pt1}. In this paper, we consider only the non-extremal cases so we always assume $M\ne 0$.\par
By (\ref{map}), we map $x_k^i, y_k^i$ and $z_k^i$ to creation operators. In order to obtain creation operators in the momentum space, we perform the mode expansion of the bosonic oscillator
\begin{align}
\label{modeExp}
\alpha^{(r)i\dagger}_k=\frac{1}{\sqrt{J_r}}\sum_{n=-\infty}^\infty e^{\frac{2\pi n k}{J_r}}a_n^{(r)i\dagger},\quad r=1,2,3
\end{align}
Let us investigate the part of spin vertex corresponding to the contractions between operators 1 and 2.
\begin{align}
V_{12}=&\prod_{k=1}^{L_{12}}(1+\alpha_{L_1-k+1}^{(1)i} \alpha_k^{(2)i})\approx\exp\left(\frac{1}{J\sqrt{r}}
\sum_{n_i^{(1)},n_i^{(2)}}\sum_{k=0}^{J_2}e^{\frac{2\pi i n_i^{(2)}k}{J_2}-\frac{2\pi i n_i^{(1)}k}{J_1}}a^{(1)i\dagger}_{n_i^{(1)}} \; a^{(2)i\dagger}_{n_i^{(2)}}   \right)\\\nonumber
=&\exp\left(-\sum_{n_i^{(1)},n_i^{(2)}}
\mathcal{N}^{12}_{n_i^{(1)}
n_i^{(2)}}\; a_{n_i^{(1)}}^{(1)i\dagger}\; a_{n_i^{(2)}}^{(2)i\dagger}    \right).
\end{align}
In the first line, the summation over $k$ gives
\begin{align}
\sum_{k=0}^{J_2}e^{\frac{2\pi i n_i^{(2)}k}{J_2}-\frac{2\pi i n_i^{(1)}k}{J_1}}\simeq J e^{-\pi i r n_i^{(1)}}\frac{\sin\pi r n_i^{(1)}}{\pi(n_i^{(1)}-n_i^{(2)}/r)}=J\sqrt{r}(-1)^{n_i^{(1)}+
n_i^{(2)}}\;  e^{-\pi i r n_i^{(1)}}\; {N}^{12}_{n_i^{(1)}n_i^{(2)}}
\end{align}
Therefore our Neumann coefficient from spin vertex is related to the Neumann coefficient in SFT \cite{Spradlin:ExplicitNeumann,Sakura:NeumannCoefficient} by a simple phase factor
\begin{align}
\label{N12}
\mathcal{N}^{12}_{n_i^{(1)}n_i^{(2)}}=(-1)^{n_i^{(1)}+n_i^{(2)}} \; e^{-\pi i r n_i^{(1)}}\; N^{12}_{n_i^{(1)}n_i^{(2)}}.
\end{align}
The explicit form of Neumann coefficient at the leading order of large $\mu$ expansion can be found in appendix A. Similarly, for the contractions between operators 1 and 3, we have
\begin{align}
V_{13}=&\prod_{k=1}^{L_{13}}(1+\alpha^{(1)i}_k \alpha_{L_3-k+1}^{(3)i})\simeq\exp\left(\frac{\sqrt{1-r}}{J}
\sum_{n_i^{(1)},n_i^{(3)}}\sum_{k=0}^{J_3} e^{\frac{2\pi i n_i^{(1)} k}{J_1}-\frac{2\pi i n_i^{(3)}k}{J_3}}\; a_{n_i^{(1)}}^{(1)i\dagger} \;
a_{n_i^{(3)}}^{(3)i\dagger}    \right)\\\nonumber
=&\exp\left(-\sum_{n_i^{(1)},n_i^{(3)}}
\mathcal{N}^{13}_{n_i^{(1)}n_i^{(3)}} \; a_{n_i^{(1)}}^{(1)i\dagger}\; a_{n_i^{(3)}}^{(3)i\dagger}  \right)
\end{align}
where our Neumann coefficient is related to the SFT Neumann coefficient by
\begin{align}
\label{N13}
\mathcal{N}_{n_i^{(1)}n_i^{(3)}}^{13}=(-1)^{n_i^{(1)}}\; e^{-i\pi r n_i^{(1)}}\; {N}^{13}_{n_i^{(1)}n_i^{(3)}}
\end{align}
For the contractions between operators 2 and 3, we have
\begin{align}
\label{V23}
V_{23}=\prod_{k=1}^{M}\alpha_k^{(3)i} \alpha_{L_2-k+1}^{(2)i}=\prod_{k=1}^M\frac{1}{J\sqrt{r(1-r)}}
\sum_{n_i^{(2)},n_i^{(3)}}e^{\frac{2\pi ik n_i^{(3)}}{J_3}-\frac{2\pi i k n_i^{(2)}}{J_2}}\; a_{n_i^{(2)}}^{ (2)i\dagger}\; a_{n_i^{(3)}}^{(3)i\dagger}
\end{align}
By dilute gas approximation $N\ll J_i$, so that in the above product $k\le N\ll J_2,J_3$. Also we assume that the momentum of the excitations are small, meaning $n_i^{(2)}\ll J_2$ and $n_i^{(3)}\ll J_3$. Therefore the phase factor in (\ref{V23}) is trivial
\begin{align}
e^{\frac{2\pi ik n_i^{(3)}}{J_3}-\frac{2\pi i k n_i^{(2)}}{J_2}}\approx 1
\end{align}
and $V_{23}$ simplifies to
\begin{align}
V_{23}=\left(\frac{1}{J\sqrt{r(1-r)}}\sum_{n_i^{(2)},n_i^{(3)}} a_{n_i^{(2)}}^{\dagger (2)i}\; a_{n_i^{(3)}}^{\dagger (3)i}  \right)^M=\left(\frac{J}{4\pi\mu |\alpha_{(1)}|}\right)^{-M}\left(-\sum_{n_i^{(2)},n_i^{(3)}}
\mathcal{N}_{n_i^{(2)}n_i^{(3)}}^{23}\; a_{n_i^{(2)}}^{\dagger (2)i}\; a_{n_i^{(3)}}^{\dagger (3)i}\right)^M.
\end{align}
where our Neumann coefficient is given by
\begin{align}
\label{N23}
\mathcal{N}_{n_i^{(2)}n_i^{(3)}}^{23}=(-1)^{n_i^{(2)}}
{N}_{n_i^{(2)}n_i^{(3)}}^{23}.
\end{align}
Let us consider the following vertex
\begin{align}
\tilde{V}_{23}=\exp\left(-\sum_{n_i^{(2)},n_i^{(3)}}
\mathcal{N}_{n_i^{(2)}n_i^{(3)}}^{23}\; a_{n_i^{(2)}}^{\dagger (2)i}\; a_{n_i^{(3)}}^{\dagger (3)i}\right).
\end{align}
which can be expanded as
\begin{align}
\tilde{V}_{23}=\sum_{M=0}^\infty\frac{1}{M!} \left(-\sum_{n_i^{(2)},n_i^{(3)}}
\mathcal{N}_{n_i^{(2)}n_i^{(3)}}^{23}\; a_{n_i^{(2)}}^{\dagger (2)i}\; a_{n_i^{(3)}}^{\dagger (3)i}\right)^M
\end{align}
For a given configuration, the number $M$ is fixed and the action of states will automatically pick out the term in $\tilde{V}_{23}$ with the corresponding $M$. Hence we can replace $V_{23}$ by
\begin{align}
\label{effective}
V_{23}\longrightarrow \left(\frac{J}{4\pi\mu |\alpha_{(1)}|}\right)^{-M} M!\cdot\tilde{V}_{23}
\end{align}
Recalling that at leading order
\begin{align}
M=\frac{1}{2}(L_2+L_3-L_1)=\frac{1}{2}(\Delta_2+\Delta_3-\Delta_1),
\end{align}
we can write the factors in front of $\tilde{V}_{23}$ in (\ref{effective}) as
\begin{align}
\left(\frac{J}{4\pi\mu |\alpha_{(1)}|}\right)^{-M} M!=\left(\frac{J}{4\pi\mu |\alpha_{(1)}|}\right)^{-(\Delta_2+\Delta_3-\Delta_1)/2}\Gamma\left(\frac{\Delta_2+\Delta_3-\Delta_1}{2}+1   \right)
\end{align}
In SFT \cite{Dobashi:3pt1}, the factor $J_1/4\pi\mu|\alpha_{(1)}|$ is the leading term of the large $\mu$ expansion of the following factor
\begin{align}
f\frac{J_2J_3}{J_1}=\frac{J}{4\pi\mu|\alpha_{(1)}|}+\mathcal{O}(\frac{1}{\mu^2})
\end{align}
so that at the leading order, we have
\begin{align}
\left(\frac{J}{4\pi\mu |\alpha_{(1)}|}\right)^{-M} M!\simeq \left(f\frac{J_2J_3}{J_1}\right)^{-(\Delta_2+\Delta_3-\Delta_1)/2}\Gamma\left(\frac{\Delta_2+\Delta_3-\Delta_1}{2}+1   \right)
\end{align}
The right hand side is nothing but the function $G(\Delta_1,\Delta_2,\Delta_3)$ in (\ref{HolRe2}) and (\ref{G}). From our derivation, it is clear that the function $G(\Delta_1,\Delta_2,\Delta_3)$ is intimately related to the interaction between the two outgoing states. Therefore it is crucial for the matching between SFT calculation and the non-extremal three-point functions.\par
We can define our spin vertex operator as
\begin{align}
V_{\text{BMN}}=G(\Delta_1,\Delta_2,\Delta_3)V_{12}V_{13}\tilde{V}_{23}.
\end{align}
This is very close to the 3-point vertex from string theory including the correct $G$-factor, except that our Neumann coefficients seem to be different from those of SFT by some phase factors. We shall show that these phase factors are trivial if we consider the physical states that satisfy the level matching conditions. When acting physical states on the spin vertex, we obtain the product of Neumann coefficients of the following type
\begin{align}
(-1)^{L_{12}+L_{23}+L_{13}}\prod_{12}
\mathcal{N}^{12}_{p_i^{(1)}p_i^{(2)}}
\prod_{13}\mathcal{N}^{13}_{q_i^{(2)}q_i^{(3)}}\prod_{23}
\mathcal{N}^{23}_{r_i^{(2)}r_i^{(3)}}\\\nonumber
=\texttt{phase}\cdot(-1)^{L_{12}+L_{23}+L_{13}}\prod_{12}
N^{12}_{p_i^{(1)}p_i^{(2)}}\prod_{13}N^{13}_{q_i^{(2)}q_i^{(3)}}
\prod_{23}N^{23}_{r_i^{(2)}r_i^{(3)}}.
\end{align}
From (\ref{N12}),(\ref{N13}) and (\ref{N23}), the phase factor is
\begin{align}
\texttt{phase}=&\prod_{12}(-1)^{p_i^{(1)}+p_i^{(2)}}e^{-\pi i r p_i^{(1)}}\prod_{13}(-1)^{q_i^{(1)}}e^{-\pi i r q_i^{(1)}}\prod_{23}(-1)^{r_i^{(2)}}\\\nonumber
=&(-1)^{\sum_i p_i^{(1)}+q_i^{(1)}}(-1)^{\sum_i p_i^{(2)}+r_i^{(2)}}e^{-\sum_i\pi i(p_i^{(1)}+q_i^{(1)})}=1
\end{align}
where we have used the level matching conditions
\begin{align}
\sum_i p_i^{(1)}+q_i^{(1)}=\sum_i p_i^{(2)}+r_i^{(2)}=0.
\end{align}
This means that if we consider the physical states, we can replace our Neumann coefficients $\mathcal{N}_{m,n}^{rs}$ by the Neumann coefficients of SFT $\tilde{N}_{m,n}^{rs}$ since the phase factors cancels out. To sum up, from the spin vertex in the BMN limit, we obtain at the leading order the following cubic vertex
\begin{align}
|V_\text{BMN}\rangle=G(\Delta_1,\Delta_2,\Delta_3)
\exp\left[-\frac{1}{2}\sum_{r,s=1\atop r\ne s}^3a_m^{(r)i\dagger}N^{rs}_{mn}a_n^{(s)i\dagger}\right]|0\rangle.
\end{align}
Notice that in the exponent we impose the condition $r\ne s$, while in SFT the Neumann coefficients $\tilde{N}^{rr}_{mn}$, which corresponds to interactions between the excitations of the same string, are non-zero. However these Neumann coefficients will appear only at higher orders in the large $\mu$ expansion. It is an interesting question whether we can obtain this kind of Neumann coefficients from weak coupling at higher loops, which we leave for future investigation. Therefore at the leading order, we have
\begin{align}
|V_{\text{BMN}}\rangle=G(\Delta_1,\Delta_2,\Delta_3)|V\rangle.
\end{align}\par
To complete our derivation, we also need to show that the spin chain states in the BMN limit also takes the same form as the ones in SFT. In the BMN limit, the scattering phases are zero which means there is no interaction between excitations. Therefore, the wave functions of the spin chain states are simply given by plane waves. In terms of bosonic oscillators, a BMN state at the leading order can be represented by
\begin{align}
\langle n_1,\cdots,n_N|=\frac{1}{\sqrt{L^N}}\sum_{x_1,\cdots,x_N=1}^L\langle0|\alpha_{x_1}^{i_i}\cdots\alpha_{x_N}^{i_N} e^{-\frac{2\pi i}{L}(n_1 x_1+\cdots+n_N x_N)}
\end{align}
where $L$ is the length of the spin chain, $\alpha_{x_k}^{i_k}$ is the bosonic oscillator introduced in (\ref{map}) and creates an excitation at position $x_k$ with polarization $i_k$ from the dual vacuum. Here $n_1,\cdots,n_N$ are the mode numbers of the excitations. Performing the mode expansion in (\ref{modeExp}), we have simply
\begin{align}
\langle n_1,\cdots,n_N|=\langle0|a_{n_1}^{i_1}\cdots a_{n_N}^{i_N}
\end{align}
which takes exactly the same form as the states in SFT.\par

There is another difference between the spin vertex and string vertex. In SFT, one has to take into account the non-trivial prefactor while we do not have similar prefactor in the spin vertex formalism. However, as argued by Dobashi and Yoneya \cite{Dobashi:3pt1}, at the leading order, the action of prefactor on the states gives rise to a simple factor, which cancels neatly the holographic factor $1/\mu(\Delta_2+\Delta_2-\Delta_1)$. Let us briefly review how this works. The prefactor for scalar excitations at the leading order reads
\begin{align}
\mathcal{P}_h=&\mu(a_0^{(2)i\dagger}a_0^{(2)i}+a_0^{(3)i\dagger}a_0^{(3)i}-a_0^{(1)i\dagger}a_0^{(1)i})\\\nonumber
+&\frac{\mu}{2}\sum_{m=1}^\infty(a_m^{(2)i\dagger}a_m^{(2)i}+a_m^{(3)i\dagger}a_m^{(3)}-a_m^{(1)i\dagger}a_m^{(1)i}+[m\rightarrow-m])\\\nonumber
+&\frac{\mu}{2}\sum_{m=1}^\infty(a_m^{(2)i\dagger}a_{-m}^{(2)i}+a_m^{(3)i\dagger}a_{-m}^{(3)i}-a_m^{(1)i\dagger}a_{-m}^{(1)i}+[m\rightarrow-m])
\end{align}
The operators in the first two lines take the form of counting operators $a_m^\dagger a_m$, hence once acted on physical states, they just count the number of difference of excitations between the incoming state 1 and out-going states 2 and 3, which is $N_2+N_3-N_1=2M$. The contribution from the first two lines is simply $\mu M$. In order to calculate the contribution from the third line, we need to make use of the symmetry of the Neumann coefficients ${N}_{-n,m}^{rs}={N}_{n,-m}^{rs}$, ${N}_{mn}^{rs}={N}_{-m,-n}^{rs}$. The operators in the third line $a_m^\dagger a_{-m}$ change the sign of mode numbers. Consider first the contractions between operators 1 and 2. Changing the sign of mode numbers of either operator gives the same result due to the symmetry of the Neumann coefficient. It is the same for the contractions between operators 1 and 3. Therefore the contributions for the impurity preserving part cancel and we only need to consider the contractions between impurity non-preserving part 2$\leftrightarrow$3. We recall the the Neumann coefficient $N_{m,n}^{23}$ at the leading order does not depend on mode number at all, hence the sign changing operators have the same effect as counting operators. The contribution is again $\mu M$. To sum up, the action of the prefactor on physical states gives rise to a multiplication of the simple factor $2\mu M$, which can be written in the following way at the leading order
\begin{align}
2\mu M=\mu(\Delta_2+\Delta_3-\Delta_1).
\end{align}
Therefore, we have

\begin{align}
\frac{1}{\mu(\Delta_2+\Delta_3-\Delta_1)}\langle1|\langle2|\langle3|H_3\rangle=\frac{1}{\mu(\Delta_2+\Delta_3-\Delta_1)}\langle1|\langle2|\langle3|\mathcal{P}_h|V\rangle=\langle1|\langle2|\langle3|V\rangle
\end{align}
Finally, gathering all the pieces, we have

\begin{align}
C_{123}=&\frac{\sqrt{J_1J_2J_3}}{N}\; \langle1|\langle2|\langle3
|V_{\text{BMN}}\rangle
=\frac{\sqrt{J_1J_2J_3}}{N}\; G(\Delta_1,\Delta_2,\Delta_3)\; \langle1|\langle2|\langle3|V\rangle=\\\nonumber
&\frac{\sqrt{J_1J_2J_3}}{N}\; \frac{G(\Delta_1,\Delta_2,\Delta_3)}
{\mu(\Delta_2+\Delta_3-\Delta_1)}\; \langle1|\langle2|\langle3|
\mathcal{P}_h|V\rangle
=\frac{G(\Delta_1,\Delta_2,\Delta_3)}
{\mu(\Delta_2+\Delta_3-\Delta_1)}\; \langle1|\langle2|\langle3|
\frac{\sqrt{J_1J_2J_3}}{N}|H_3\rangle
\end{align}
where the first line is from the \emph{spin vertex} while the last line is exactly the holographic relation (\ref{HolRe2}) proposal by Dobashi and Yoneya. Therefore we have shown that gauge theory computation reproduces exactly SFT result at tree level.

\section{Conclusion and Outlook}

In this paper, we performed the BMN limit of the spin vertex at leading order of $\lambda'$ for the scalar excitations and showed that it matches the string vertex in SFT. The match between CFT and SFT computations at tree level was proved in \cite{Dobashi:3pt1}. Our aim here was to obtain this match using the spin vertex approach. In this way the exponential form of the spin vertex appears naturally, and allows to introduce the Neumann coefficients from the CFT side of the duality. The string vertex consists of three parts at any order in the expansion parameter: the exponential with the Neumann coefficients, the prefactor ${\cal P}_h$ and the G-factor. We hope the same structure holds for the spin vertex in BMN limit at any loop. When taking into account the radiative corrections there are two parts which receive corrections: the Neumann coefficients and the G-factor. The Neumann coefficients for pp-wave string vertex were computed in \cite{Spradlin:ExplicitNeumann} and \cite{Sakura:NeumannCoefficient} to all orders in $1/\mu^2$. If the structure of the spin vertex holds at any order of coupling constant and one manage to establish the connection between the CFT and the SFT Neumann coefficients at any loop, then we can reproduce exponential part of the spin vertex at any coupling. Another problem is the G-factor. As it has been already mentioned, its expansion in coupling constant gives rise to $\log\mu$ dependent terms and it's not clear how to deal with them. Fixing the G-factor at any loop order from CFT and SFT sides, together with all-loop expression for Neumann coefficients, would provide us with the all-loop result for three-point functions in BMN limit, at least in the compact sector.

Another interesting direction to explore is to extend this approach to other sectors, that is to include vector and fermionic excitations. The main problem is that in general in $\mathcal{N}=4$ SYM the space-time dependence of three-point function is more complicated than in the compact sector and there are more than one structure constants to be fixed. On the contrary, the SFT calculation provides us with only one quantity. It is important to understand the relation between it and the structure constants from the gauge theory side. This problem was discussed in the literature for vector \cite{Dobashi:3pt1} as well as for fermionic \cite{Dobashi:fermions}, \cite{Georgiou:fermions} excitations, but the full understanding is still missing.

Very recently, the authors of \cite{BJ} reinterpreted the Neumann coefficients as a kind of generalized form factor. They found that the Neumann coefficients satisfy a set of functional equations similar to the axioms of form factor bootsrap equations. This observation provides us a novel point of view to understand and construct Neumann coefficients.

\appendix
\section{Large $\mu$ behavior of Neumann coefficients}
In this appendix, we list the leading order of Neumann coefficients in the large $\mu$ expansion. We take the same convention as in \cite{Dobashi:3pt1}. For $(m,n)\ne(0,0)$
\begin{align}
&N_{mn}^{22}=\frac{(-1)^{m+n}}{4\pi\mu|\alpha_{(1)}|r},\quad N_{mn}^{23}=\frac{(-1)^{m+1}}{4\pi\mu|\alpha_{(1)}|\sqrt{r(1-r)}}\\\nonumber
&N_{mn}^{33}=\frac{1}{4\pi\mu|\alpha_{(1)}|(1-r)},\quad N_{mn}^{11}=\frac{(-1)^{m+n+1}\sin(\pi m r)\sin(\pi n r)}{\pi\mu|\alpha_{(1)}|}\\\nonumber
&N_{mn}^{21}=\frac{(-1)^{m+n+1}\sin\pi n r}{\pi\sqrt{r}(n-m/r)},\quad N_{mn}^{31}=\frac{(-1)^n\sin(\pi n r)}{\pi\sqrt{1-r}(n-m/(1-r))}.
\end{align}
For $(m,n)=(0,0)$, we have
\begin{align}
&N_{00}^{11}=0,\quad N_{00}^{12}=-\sqrt{r},\quad N_{00}^{13}=-\sqrt{1-r}\\\nonumber
&N_{00}^{23}=-\frac{1}{4\pi\mu|\alpha_{(1)}|\sqrt{r(1-r)}},\quad N_{00}^{22}=\frac{1}{4\pi\mu|\alpha_{(1)}r},\quad N_{00}^{33}=\frac{1}{4\pi\mu|\alpha_{(1)}|(1-r)}
\end{align}

\acknowledgments
It is our pleasure to thank Ivan Kostov and Didina Serban for guidance and many useful discussions. We are also indebted to Romuald Janik for pointing to us this problem and sharing his knowledge on the subject. Finally, we thank Zoltan Bajnok and Shota Komatsu for their interest on the current work. This work has
received support from the People
Programme (Marie Curie Actions) of the European Union's Seventh
Framework Programme FP7/2007-2013/ under REA Grant Agreement No
317089 (GATIS).

\bibliographystyle{JHEP}
\providecommand{\href}[2]{#2}\begingroup\raggedright\endgroup

\end{document}